# On Using Blockchains for Beyond Visual Line of Sight (BVLOS) Drones Operation: An Architectural Study

TAHINA RALITERA, Université Paris-Saclay, CEA, List, France
AGNES LANUSSE, Université Paris-Saclay, CEA, List, France
ÖNDER GÜRCAN, Université Paris-Saclay, CEA, List, France

Beyond Visual Line of Sight operation enables drones to surpass the limits imposed by the reach and constraints of their operator's eyes. It extends their range and, as such, productivity, and profitability. Drones operating BVLOS include a variety of highly sensitive assets and information that could be subject to unintentional or intentional security vulnerabilities. As a solution, blockchain-based services could enable secure and trustworthy exchange and storage of related data. They also allow for traceability of exchanges and perform synchronization with other nodes in the network. However, most of the blockchain-based approaches focus on the network and the protocol aspects of drone systems. Few studies focus on the architectural level of on-chip compute platforms of drones. Based on this observation, the contribution of this paper is twofold: (1) a generic blockchain-based service architecture for on-chip compute platforms of drones, and (2) a concrete example realization of the proposed generic architecture.

## 1 INTRODUCTION

International Civil Aviation Organization (ICAO) defines BVLOS operation as "an operation in which the remote pilot or Remotely Piloted Aircraft (RPA) observer does not use visual reference to the remotely piloted aircraft in the conduct of flight" [13]. This type of operation is crucial for the large-scale expansion and economic feasibility of a large range of drone business applications such as logistics, mapping, surveying, etc. BVLOS operation enables drones to surpass the limits imposed by the reach and constraints of their operator's eyes, extending their range and, as such, productivity and profitability.

However, drones include a variety of highly sensitive assets and information that could be subject to unintentional or intentional security vulnerabilities. Attacks may arise from a number of sources, including malicious intellectual property blocks (IPs) in the hardware, malicious or vulnerable firmware and software, insecure communication of the system with other devices, eavesdropping or "man-in-the-middle" attacks, breach of confidentiality of the stored data, corruption of the integrity of collected data [20]. Therefore, secure, resilient, and tamper-proof storage is important for security and accountability of the data.

Based on this observation, we propose a blockchain-based solution for BVLOS drones operation. Indeed, the blockchain technology has the potential to secure data that is being dynamically updated, through its security capabilities such as hashing, smart contracts, consensus protocols, public and private keys, etc. In this way, the data is stored in an immutable manner, making it impossible to modify neither intentionally nor unintentionally. The blockchain also allows for traceability of exchanges and performs synchronization with other nodes in the network. Therefore, blockchain-based services could enable trustworthy exchange and storage of the drones related data. Moreover, it is ideally suited for drone applications which are very dynamic in nature [1].

Drones can be used for various purposes and use cases. However, using a concrete blockchain-based solution with well-defined transaction types and parameters may be complicated for these different purposes and use cases. Consequently, we identify the generic component necessary at the architectural level and propose a solution at that level. Concretely, the contributions of this paper are as follows:

- A generic blockchain-based service architecture dedicated to BVLOS drones operation.
- A concrete example realization of the proposed generic architecture.

The remaining of this paper is organized as follows. Section 2 describes the background on blockchain. Section 3 describes our proposed generic blockchain-based service architecture for drones operating BVLOS. Section 4 shows the effectiveness of the proposed architecture on a realistic case study. Section 5 gives related work and Section 6 concludes the paper.

## 2 BACKGROUND

Blockchain systems [18] are composed of participants communicating over a **peer-to-peer network** using transactions. Its overall objective is to maintain the consistency and coherency of all transactions in a replicated immutable transparent append-only **data structure called the blockchain**.

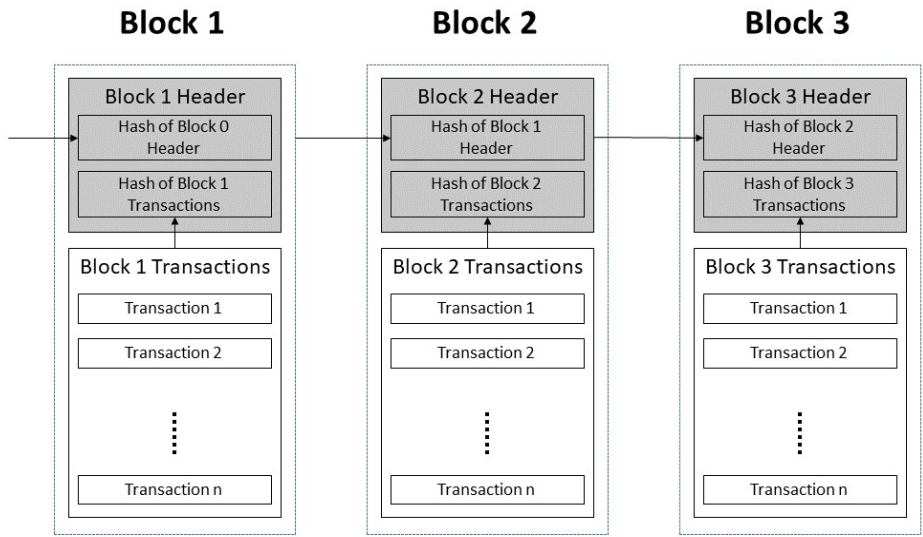

Fig. 1. Schematic View on the Blockchain Data-Structure

As illustrated in Figure 1, this data structure consist of blocks which are connected to each other using cryptographic links. Each block contains an ordered list of transactions. Transactions usually consists of references to accounts and formal elements like the type of the transaction, its ID, as well as cryptographic elements like hashes and signatures.

Accounts on the blockchain serve as senders and receivers of transactions. An account consists of a public key and the corresponding private key, which together form an account key pair. The public key is used by senders and receivers of transactions while the private key is used to create digital signatures to authorize transactions.

Participants of a blockchain system can play different roles in a blockchain network. While some can be responsible for issuing transactions, others can be responsible for creating the blocks and having consensus on the created blocks[1].

Similarly, storage may also differ. While some participants store the full replica of the blockchain (i.e. full nodes) [8], some others may choose to store less information due to their needs and capabilities (i.e. light nodes) [21].

## 3 GENERIC BLOCKCHAIN-BASED SERVICE ARCHITECTURE

Blockchain-based services consist of subsystems which use a blockchain-based solution integrated with the communication architecture of the BVLOS drone operation. The objective is to design a blockchain solution deployable to the drones and the Ground Stations (GS) computing platforms. This will allow adding services (authentication, trust management) using trustworthy data securely.

For that purpose, we model the entities of the U-Space (i.e. drones and ground stations) as communicating network nodes where a local replicate of the shared blockchain is maintained. The proposed solution is intended to be embedded inside drones and ground stations for securing the sensitive data used by the on-chip compute platforms in the future.

Application is made in the context of the Airborne Data Collection on Resilient System Architectures (ADACORSA) European project [19]. Figure 2 shows the blockchain-based services (LS_Blcockchain_based_services) positioning within the overall ADACORSA system architecture where secure communications, drone services and external entities (i.e., drones and ground stations) are represented as Logical Actors. Note that these components are not currently embedded in real drones but are positioned like this for evaluation purposes. In the future real implementations, they would be integrated at the Reliable Communication component level, so most of the communications are handled through the communication layer.

The blockchain-based services system consists of a main component called Tamper-proof Storage Service (TSS) and dedicated service components embedded in the drones and the ground stations. Thanks to TSS, the drones and the ground stations collaboratively maintain and share sensitive data in such a way that it is traceable, and that its integrity is protected from intentional and unintentional modifications. Blockchain technology is used to store, distribute and validate data using a chain of cryptographically connected blocks. Concretely, the blockchain peer-to-peer network consists of drones and grounds stations and the blockchain data structure is stored and managed by TSS.

TSS is developed to provide a common abstract view and interface to permit in depth study of characteristics and performances of blockchain solutions. It externalizes the interactions with the blockchain by making explicit interactions and transactions management. Concretely, it interacts with other services using interfaces for transaction submission. In addition, internal management services are provided for the synchronization of nodes and for protocol support. As illustrated in Figure 3, TSS is composed of two components: Blockchain Management and Local Blockchain Storage.

*3.0.1 Blockchain Management.* Blockchain Management is responsible for generic blockchain operations such as registering data into the Local Blockchain storage, verifying the validity of the data, reading stored data and synchronizing data within the network. In addition, this sub-component is responsible for the management of the external services (Service A and Service B) related data structures and operations. It also contains dedicated APIs and libraries, as well as internal management and applicative transactions.

---

[1]Such consensus can be achieved using various algorithms. For a review, see [22].

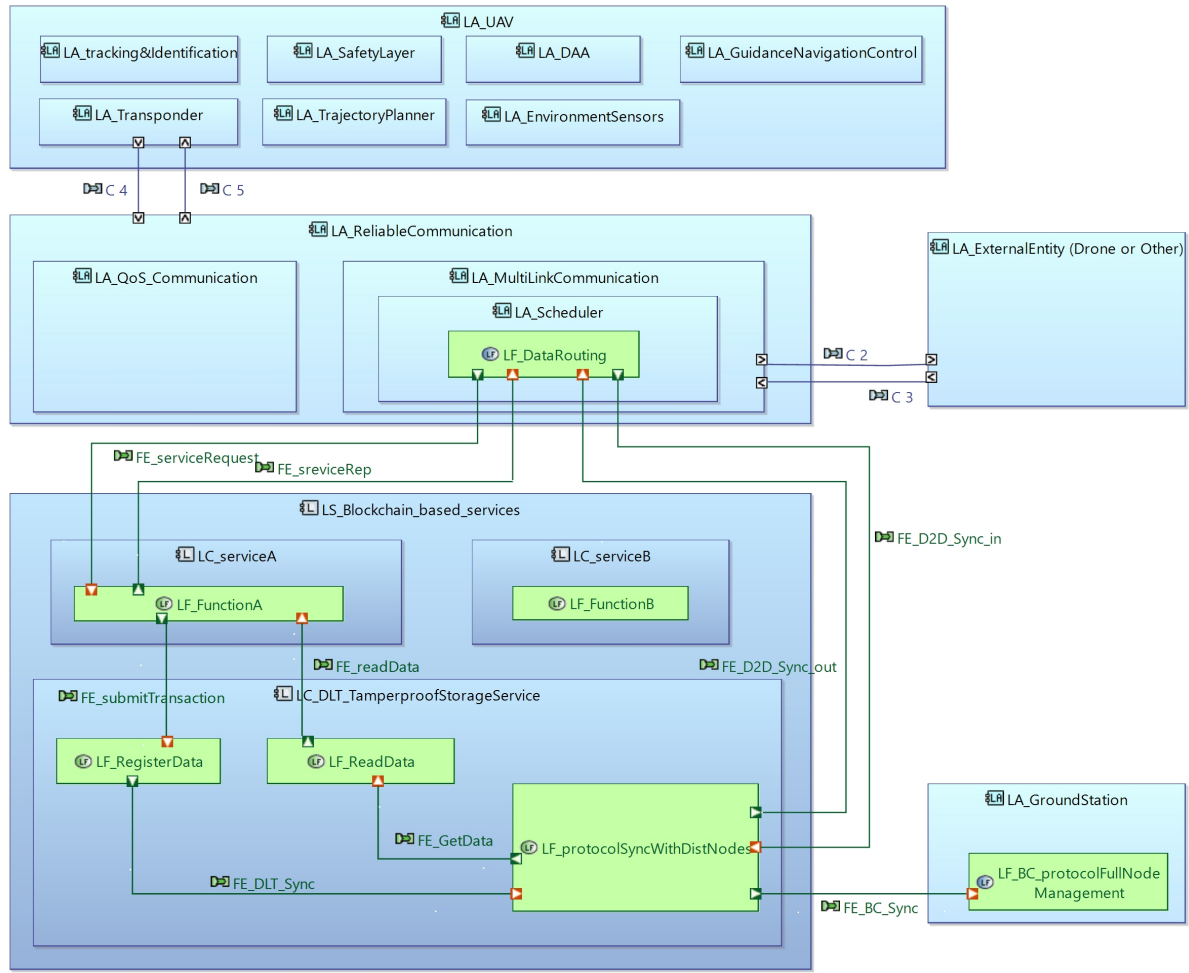

Fig. 2. Block Diagram within the Overall System Architecture

Basically, there are two categories of transaction types that are supported:

- default blockchain transactions such as coin base transaction and token transfer transaction
- applicative transactions dedicated to specific services. Examples are given in 4.

3.0.2 *Blockchain Storage.* The Local Blockchain Storage is responsible for storing data in the form of a time-stamped network of secure logs of data. Concretely, to provide services in a trustworthy and decentralised way, the blockchain-based services can require the data stored in the Local Blockchain Storage.

Due to the memory and computation requirement of TSS, it is not necessary that every device maintains a full replica of the blockchain locally. Depending on the needs and capabilities of the support devices, the TSS can be realized in two ways: as a lightweight node or as a full node. In the example illustrated in 4 "Blockchain-based Authentication Architecture", full nodes are deployed on ground stations while light nodes are deployed on drones. Thus, TSS for drones communicates and relies on full nodes (i.e. TSS for Ground Station) to provide them with the necessary information.

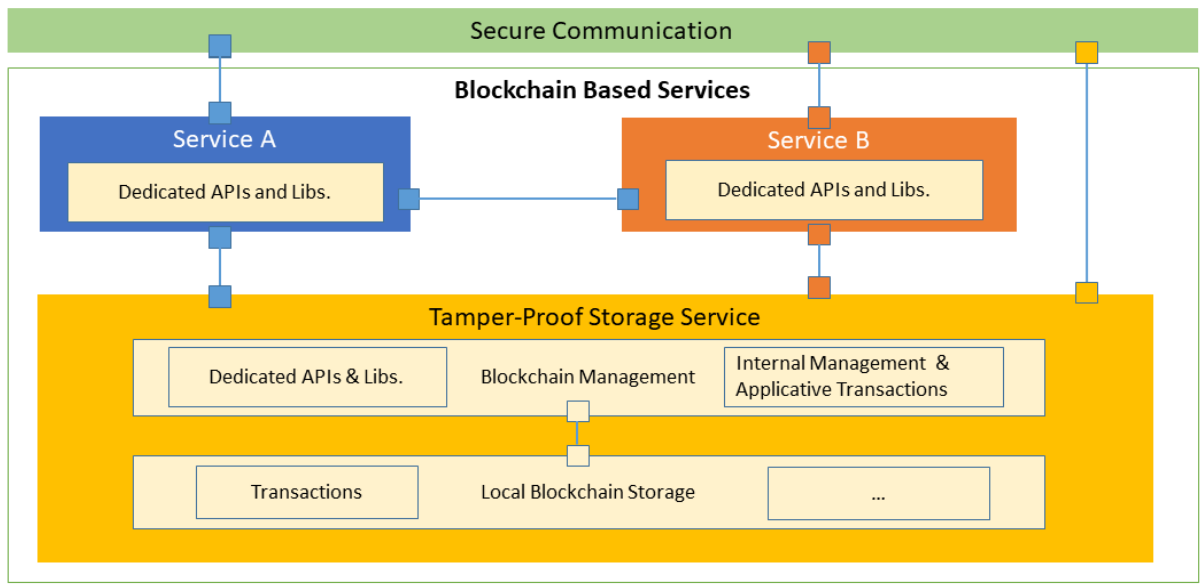

Fig. 3. Block Diagram of Blockchain-Based Services

## 4 CASE STUDY: BLOCKCHAIN-BASED AUTHENTICATION ARCHITECTURE

This section provides a concrete realization of the generic architecture described previously in section 3 : a blockchain-based authentication service for BVLOS drones operation.

Providing decentralised authentication functionalities contributes to secure communication and trustworthy cooperation in open networks of (semi-)autonomous drones and the Internet of Things. The Authentication subsystem (LC_Authentication illustrated in Figure 4) enables drones operating BVLOS to authenticate other entities in the network that can be used subsequently to establish secure communication channels between them. The primary focus is on open networks of drones such as flying ad-hoc networks, but the developed solutions may also be suitable for networks containing other entities like cars, smart sensors, etc. The main functions provided are decentralised entity authentication in networks of drones and secure storage of authentication-related data.

The system consists of two main components :

- TSS component (LC_DLT_TamperProofStorageService), previously described in section 3. In this context, TSS is responsible for secure, permanent storage and management of the authentication data in a decentralised way. Thanks to TSS, the drones and the Ground Stations collaboratively maintain and share sensitive data so that nobody can alter the stored data neither intentionally nor unintentionally.
- Authentication Service (AS) component (LC_Authentication) which provides the protocol and cryptographic operations required for authentication. It also handles authentication requests. AS provides this service in a decentralised way by using TSS. Concretely, for its operation it requires the data stored in the TSS component.

AS component provides its service (i.e. entity authentication) to cooperative applications between drones. It uses the reliable communication hardware and software of the drone to communicate with other drones. The drones use the authentication protocol provided by the AS component in the process of establishing secure communication channels to authenticate the communication partner. In the scope of this paper, the AS component abstracts from the details of the communication stack in accordance with the OSI model.

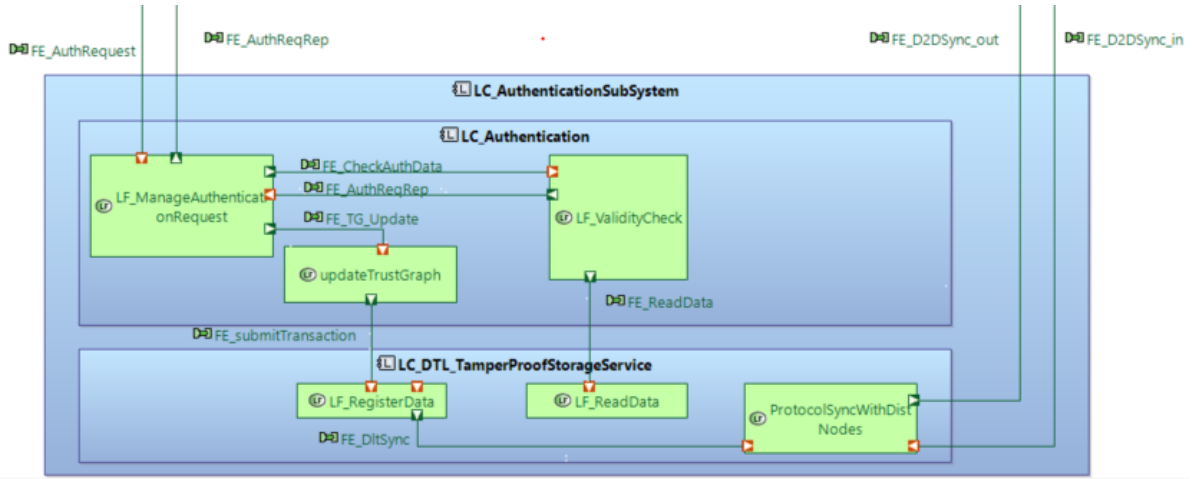

Fig. 4. Block Diagram of the Blockchain-Based Authentication Service Architecture

To ensure decentralised resilient and tamper-proof properties to the solution, it uses the so-called TSS component that uses blockchain-based technologies notably for storage and consistency. TSS brings by construction security properties such as verifiable interactions, resilience, and accountability. Due to the distributed nature of TSS, it is not necessary that every drone maintains such a component in principle. However, it is assumed that every drone maintains its own TSS component in order to provide the service of authentication in cases where a drone does not have access to a TSS component at a trusted ground station.

### 4.1 Authentication Service (AS)

The task of this component is to authenticate drones and other entities in networks of drones, operating BVLOS, e.g., in Flying Ad-hoc NETworks (FANETs). The authentication service outlined in this paper is based on public key cryptography. The secure usage of public key cryptography is enabled by a blockchain-based Public Key Infrastructure (PKI) described in details in [14]. Its operation can be summarised as follows:

The PKI binds identities to public keys whereby the public keys can be used in a secure manner during the authentication process. The blockchain-based PKI is used to store the identities together with their public keys and the confirmations of bonds between identity and public keys. Concretely, the task of a PKI is to guarantee the authenticity of the public keys, i.e. a certain public key belongs to a certain identity. This allows the secure usage of public key cryptography.

In the case an entity wants to use the public key of a not previously known identity, it has to ensure that the public key is controlled by the stated identity. For this purpose, the entity requires a confirmation of this bond between identity and public key by a trusted source. Confirmations can have different forms, e.g., digital certificates, or in the proposed system special transaction of the blockchain system. These confirmations establish (trust) relationships between the entities.

Mathematically, the authors of [14] model these relationships as a directed graph, called trust graph. The identities and their public keys are represented by nodes. Confirmations are represented by directed edges between the nodes. The so-called trust model defines how confirmations are interpreted, how "chains" of confirmations are evaluated and who is allowed to create confirmations.

Therefore, the task of the dedicated blockchain of the PKI is to store, update and provide the trust graph, including the public keys and identities.

For that purpose, in addition to the default transaction types (i.e., coinbase transaction and token-transfer transaction), the system uses the following applicative transactions for authentication:

- *Entity transaction*: This transaction is used to store the public key for authentication with a signature and the identity information of an entity and creates a node in the trust graph. The identity information contains the name of the entity and some characteristic properties, like the type of the entity. The balance of the account is reduced by the fee, and a certain number of tokens is reserved. Every account can only make at most one entity transaction.
- *Revoke entity transaction*: This transaction can only be made by the account if it has issued an entity transaction, previously. This transaction destroys the node and all incoming and outgoing edges in the trust graph. The fee of the transaction is paid by the reserved tokens, and the remaining is added to the balance of the account.
- *Confirmation transaction*: This transaction is used by a sender to confirm the identity-key binding of a receiver and, hence, creates a directed edge from the sender to the receiver in the trust graph and assigns the maximal allowed path length to it. The balance of the account is reduced by the fee and some tokens are reserved. If an edge between two nodes already exists, the number is updated.
- *Revocation transaction*: This transaction is used by a sender to revoke a confirmation transaction, performed by itself, and, consequently, the edge in the trust graph is destroyed. The reserved number of tokens is used to pay the fee.

Since the drones have only limited capabilities to store and process data, it is not feasible that they participate in the consensus process or store the whole blockchain. Besides their own key pairs, they require only the trust graph, i.e., the nodes and edges, public keys and identity information. They do not need the whole trust graph but only the parts that are relevant for them, i.e., they only need the data of the entities to which they are connected via a valid path. Furthermore, they may also store the block headers for the possibility to update the trust graph during a mission and to use (unsecure) channels to the blockchain network. Consequently, the node and edges are stored in an appropriate form.

Therefore, it is assumed that each drone maintains its own TSS component to provide a blockchain-based authentication service in cases where a drone does not have access to a TSS component at a ground station. TSS for drones takes the form of a lightweight node. This means that it does not store the entire copy of the chain but only queries the current state to determine the last block, and broadcasts transactions for processing. Thus, TSS for drones communicates and relies on full nodes (TSS for Ground Station) to provide them with the necessary information. TSS for Ground Station takes the form of a full node. This means that it stores a copy of the entire blockchain. Its main tasks also include maintaining the consensus between other nodes and verification of transactions.

## 5 RELATED WORK

Blockchain technology enables decentralised and trustworthy exchange and storage of drone system data. It also allows for traceability of exchanges and performs synchronization with other nodes in the network.

In this context, according to our observation, most of the interest is on blockchain-based solutions focusing on the network and communications aspects of drone systems [2, 3, 5, 7, 9]. For example, [5] presents a blockchain-based network model for unauthorized drone detection in Internet of Drone (IoD) environment. In their approach, the blockchain technology is used to record various events in the IoD environment that facilitates any AI-enabled Big data analytics to read this data from the blockchain and provide various statistical analytics to improve the system. Another approach for blockchain-based IoD is also described in [2]. The solution is called UTM-Chain. It allows to secure drone flight plan and to store drone's flight data records. In their architecture, the users, drones, and ground stations act as nodes, storing the whole blockchain and participating in the consensus protocol to verify blocks. [3] proposes a system architecture of blockchain-assisted 5G drone network where data processing and storage are handled at the edge or the cloud. Blockchain authenticates communications among drones by tracking all their transactions, makes it available to all network nodes, and achieves data integrity (i.e., cryptography) to provide tamper resistance. In [9], the authors designed a drone and blockchain-based system for Industry 4.0 inventory and traceability applications. The proposed solution consists of a communication architecture that uses a blockchain to store certain inventory data collected by drones. The ground station makes use of a software module that acts as a blockchain client. The authors of [7] also proposed a blockchain-based general architecture for data collection and management of components in drones.

While the above examples deal with network and communication aspects, several contributions have also been made on protocols and algorithms [4, 10, 15, 17]. The authors of [17] proposed a blockchain-based solution to achieve cross-domain authentication for 5G-enabled drones. Their approach employs multiple signatures based on threshold sharing to build an identity federation for collaborative domains. Reliable communication between cross-domain devices is achieved by utilizing smart contract for authentication. A blockchain-based technique to support multi-party authentication to facilitate trustworthy communications between group of drones is proposed in [10]. In [4], the authors proposed a lightweight blockchain based model to provide distributed authentication and anonymous authorization in IoD. The authors of [15] proposed a decentralised architecture of flying ad hoc nodes based on blockchain and using Practical byzantine fault tolerance (PBFT) for consensus among nodes.

In the end, few studies focus on the architectural level of on-chip compute platforms of drones. An example is described in [6], where the authors proposed a concept based on blockchain embedded in the drones and some nodes on the ground stations. Their solution is composed of two main block components:

- A Robot Operating System (ROS) middleware block in which the main drone functionalities are implemented and which integrates at least three nodes:
  - The Navigation node or drone state estimation node that provides the drone position and velocities, which are necessary to the other nodes;
  - The Control node or automatic pilot node that manages the behavior of the drone according to information received from the Navigation node and and the Blockchain Management node;
  - The Blockchain management node that makes the link between the blockchain and the other nodes.
- A blockchain block that is assumed to communicate with ROS nodes through a web API (REST/RPC). This block implements all the blockchain functionalities (e.g., synchronization with the other nodes, blockchain network management, insertions of new transactions, execution of smart contracts).

They assume that the drones only communicate together through the blockchain. The synchronization of the blockchain clients allows to transfer the written data. They also identified the Tangle-based IOTA blockchain as the best candidate among the current blockchains.

While most approaches tackle with network (or communication) and protocol (or algorithms) aspects, few studies focus on the architectural level of on-chip compute platforms of drones. Therefore, there is a lack of proposal on ligtweight blockchain-based service architecture for on-chip compute platforms of drones. Our proposed solution, on the other hand, deals with a lightweight generic blockchain-based service architecture for on-chip compute platforms of drones.

## 6 CONCLUSION AND FUTURE WORK

In this paper, we provided, in a first step, a generic blockchain-based service architecture. This solution consists of a main component called Tamper-proof Storage Service (TSS) and dedicated service component embedded on drone and ground station on-chip compute platform. TSS allows the drones and the ground stations to collaboratively maintain and share sensitive data in such a way that it is traceable, and that its integrity is protected from intentional and unintentional modifications. For that purpose, TSS contains components that are responsible for blockchain management and the local blockchain storage.

In a second step, we provided a concrete realization of this generic blockchain-based service architecture which is a blockchain-based authentication service. It enables secure communication and trustworthy cooperation in open networks of (semi-)autonomous drones and the Internet of Things. This blockchain-based authentication service is composed of an Authentication Service (AS) that provides the protocol and cryptographic operations required for authentication and a TSS that is used by the AS for secure, permanent storage and management of the authentication data in a decentralised way.

As the proposed solution is intended to be embedded inside drones and ground stations for securing the sensitive data used by the on-chip compute platforms, simulations will permit to evaluate different protocols and algorithms for future implementation on real drones and ground stations. To this end, and in the continuation of this work, a UTM blockchain simulation environment model is being developed. This model will be used for tests and evaluation based on scenarios parametrized with different types and number of drones. For that purpose, implementations are performed using an agent-based modeling and simulation framework dedicated to blockchains [11, 12, 16].

This simulation model will also allow to evaluate the performance of the suggested architecture based on the following metrics :

- Metrics on embeddability:
  - Running time
  - Energy consumption
  - Latency/Delay
- Metrics on authentication:
  - Probability of authentication
  - Degree of Reliability of the authentication

## 7 ACKNOWLEDGMENTS

This work is supported by ECSEL Joint Undertaking (JU) through the Project ADACORSA under grant agreement No 876019. The JU receives support from the European Union's Horizon 2020 research and innovation programme and Germany, Netherlands, Austria, Romania, France, Sweden, Cyprus, Greece, Lithuania, Portugal, Italy, Finland, Turkey.

The authors also thank Nicholas Jäger [14] who collaborates with us on the presented case study focusing on the algorithmic details.